# Near-locality of exchange and correlation density functionals for 1- and 2-electron systems

Jianwei Sun[1], John P. Perdew[1,2], Zenghui Yang[1], and Haowei Peng[1]

[1] Department of Physics, Temple University, Philadelphia, Pennsylvania 19122

[2] Department of Chemistry, Temple University, Philadelphia, Pennsylvania 19122

**Abstract: The uniform electron gas and the hydrogen atom play fundamental roles in condensed matter physics and quantum chemistry. The former has an infinite number of electrons uniformly distributed over the neutralizing positively-charged background, and the latter only one electron bound to the proton. The uniform electron gas was used to derive the local spin density approximation (LSDA) to the exchange-correlation functional that undergirds the development of the Kohn-Sham density functional theory. We show here that the ground-state exchange-correlation energies of the hydrogen atom and many other 1- and 2-electron systems are modeled surprisingly well by a different local spin density approximation (LSDA0). Our LSDA0 is constructed to satisfy exact constraints, but agrees surprisingly well with the exact results for a uniform two-electron density in a finite, curved three-dimensional space. We also apply LSDA0 to excited or noded 1-electron densities. Furthermore, we show that the locality of an orbital can be measured by the ratio between the exact exchange energy and its optimal lower bound.**

The uniform electron gas (UEG) and the hydrogen atom are two of the most important models in condensed matter physics and quantum chemistry. These models represent two opposite limits from several perspectives (e.g., extended vs. confined, and ∞ vs. 1 in electron number). When the density functional theory (DFT)[1-3] was developed, the UEG was first used to derive the local spin density approximation (LSDA) [4-6] to its exchange-correlation energy, the only part that needs to be approximated in DFT. LSDA was at first believed to be too crude for any practical applications, but it performed surprisingly well for solids, as later explained by the fact that LSDA satisfies exact constraints on the exchange and correlation holes[7]. However, LSDA predicts a too-high energy for H, and DFT was seldom used in chemistry until the advent of generalized gradient approximations (GGA) [8-14] which use the density gradient to account for the inhomogeneity of real electron densities and lower the energy for H.

Unfortunately, GGAs recovering the UEG limit and delivering reasonable energies for H and other atoms strongly violate a recent exact constraint (defined below as Eq (2)) for 1- and 2-electron systems[15]. This letter shows that the exact constraint implies near-locality of the exchange-correlation energies of 1- and 2-electron systems, which can be modeled by a local spin density approximation (denoted as LSDA0 to differentiate it from LSDA of UEG). Here we define a local spin density functional as one whose energy density at a point $\vec{r}$ in space depends only upon the local spin densities $n_\uparrow(\vec{r})$ and $n_\downarrow(\vec{r})$ at that point. This is the original definition, which we prefer. The unification of UEG and H as well as satisfaction of the exact constraint on 1- and 2-electron systems can all be achieved at the level of a meta-generalized gradient approximation[16], where the introduced kinetic energy density helps to recognize and treat with different GGAs these two limits[17], UEG and H.

Let's focus on the exchange energy first. The exchange energy of any spin-unpolarized density within the local density approximation (LDA) is $E_x^{LDA}[n] = \int d^3r n \varepsilon_x^{unif}(n)$, where $\varepsilon_x^{unif}(n) = -(3/4\pi)(3\pi^2 n)^{1/3}$ is the exchange energy per particle of a UEG with density $n$. The spin-scaling relation[18] can be used to extend this formula to LSDA, introducing factor of $2^{1/3}$ for a fully spin-polarized density.

In a recent Letter[19], Loos and Gill presented exact solutions for the ground-state energy of two Coulomb-interacting electrons with uniform density in a finite curved three-dimensional space at special space-curvatures $R^{-1}$. The Loos-Gill system is two electrons with uniform spin-unpolarized density in "3-spherium", the three-dimensional surface of a four-dimensional ball of radius $R$, with volume $2\pi^2 R^3$. It is a generalization of the more easily imagined 2-spherium, the two-dimensional surface of a three-dimensional ball of radius $R$, with area $4\pi R^2$. The Coulomb interaction in $D$-spherium is taken to be $1/u$, where $u \leq 2R$ is the distance between electrons measured in the flat $(D+1)$-dimensional space.

Eq. (3) of their paper[20] says that the total energy is $E = E_0/R^2 + E_1/R + E_2 + ....$ After Eq. (28) of Ref. 20, we find $E_0 = 0$. Eq. (27) of Ref. 20 gives $E_1 = 0.8488$. Now $0.8488/R = U[n] + E_x[n]$, where $U[n]$ is the Hartree electrostatic interaction of the density with itself. For a two-electron ground state, $E_x[n] = -U[n]/2$, so the exact exchange energy for the Loos-Gill system is $E_x[n] = -0.8488/R$. The UEG-based local density approximation for this system is trivially $E_x^{LDA}[n] = -0.6882/R$, so

$$E_x[n]/E_x^{LDA}[n] = 1.233. \tag{1}$$

Recently some of us[15] have used theorems of Lieb and Oxford[21,22] to prove rigorously that, for any two-electron spin-unpolarized ground state in flat infinite three-dimensional space,

$$E_x[n]/E_x^{LDA}[n] \leq 1.174. \tag{2}$$

The bound in Eq. (2) is optimal; it cannot be improved. While Eq. (1) violates the exact constraint of Eq. (2), the violation is small. So the artifacts of the Loos-Gill model are small in two-electron ground states. These artifacts should disappear in the limit $R \to \infty$ and electron number $\sim R^3$, where the model should approach the standard uniform electron gas.

Loos and Gill propose their model system as the basis for a "generalized local density approximation" (as defined in Ref. 23) that is better than UEG-based functionals for real (confined) two-electron densities in infinite flat three-dimensional space (e.g., the He atom or the $H_2$ molecule). Although motivated and constructed in a different way, our LSDA0 supports their proposal and demonstrates that the exchange-correlation energies of the two-electron ground states are indeed remarkably "local". By applying the uniform density scaling, the LSDA0 exchange energy of any spin-unpolarized density is $E_x^{LSDA0}[n] = \int d^3r n \varepsilon_x^{unif}(n) F_x$. Again the spin-scaling[18] can be used to obtain the exchange energies of spin-polarized densities. For spin-unpolarized densities, $F_x = 1.16588$, a value very close to its bound 1.174, is then determined from the exact exchange energy for the fully spin-polarized H atom. This shows that the inequality in Eq. (2) is typically close to being an equality.

The correlation part of LSDA0 can be written as $E_c^{LSDA0} = \int d^3r n\, \varepsilon_c^{LSDA0}(r_s, \zeta)$ with

$$\varepsilon_c^{LSDA0} = \frac{-b_{1c}}{1+b_{2c}r_s^{\frac{1}{2}}+b_{3c}r_s} g_c(\zeta). \tag{3}$$

Here, $\zeta = (n_\uparrow - n_\downarrow)/(n_\uparrow + n_\downarrow)$ is the relative spin polarization defined with the spin densities $n_\uparrow$ and $n_\downarrow$, and $r_s = (4\pi n/3)^{-1/3}$ is the Seitz radius. $b_{1c} = 0.0233504$ is determined by the correlation energy $E_c = -0.0467$ Ha of the high-density limit of the two-electron ion with proton number $Z \to \infty$ ( the high-density limit for helium-like ions[24]). $b_{3c} = 0.102582$ is determined by the lower bound on the exchange-correlation energies of 2-electron systems[21], which says that the exchange-correlation energy of any 2-electron density is no more negative than 1.67082 times that evaluated with the LDA exchange $E_x^{LDA}[n]$. $b_{2c} = 0.1018$ is fixed by $E_{xc}$ (He) = -1.068 Ha, the exchange-correlation energy of He[11,25]. $g_c(\zeta)$ is a function[16] that is zero for $\zeta = \pm 1$ to satisfy the one-electron self-correlation-free constraint, and one for $\zeta = 0$, but is otherwise arbitrary and meaningless for the current discussion. LSDA0 is reminiscent of the Wigner interpolation[26,27] for the correlation energy of the UEG, but is designed for finite systems. Fig. 1 plots the LSDA0 and LSDA correlation energies as functions of $r_s$. LSDA0 significantly reduces the magnitude of the correlation energy per electron in comparison with LSDA. LSDA0 doesn't have the $r_s \to 0$ logarithmic singularity of LSDA. Both features are required for finite systems.

The correlation energy of the Loos-Gill system in the limit $R \to 0$ is -0.0476 Ha[19], between those of the high-density limits for helium-like ions (-0.0467) and the Hooke's-law two-electron atoms (-0.0497). The correlation energies of LSDA0 and the exact values from Table II of Ref. 19 are respectively -0.0343 and -0.0368 Ha at $R = 1.58$ Bohr, and -0.0065 and -0.0062 at $R = 39.7$ Bohr. This comparison suggests that a local density correlation energy functional based on the Loos-Gill system would also be realistic for the correlation of the two-electron ground states.

Now let's turn to the Hooke's atoms with 2 electrons. The Hooke's atoms are model systems where the Coulombic electrons are confined by external harmonic potentials with a characteristic frequency $\omega$. The 2-electron Hooke's atoms are solvable analytically for a particular, denumerably infinite set of frequencies, and their ground-state exchange and correlation energies are known[28]. Figure 2 shows the percentage errors of $E_x$, $E_c$, and $E_{xc}$ for different functionals as a function of different frequencies or classical distances $r_0$ between electrons in the Hooke's atom. These are commonly-used semilocal density functionals, including LSDA[5], the Perdew-Burke-Ernzehof (PBE) [13] and Becke-Lee-Yang-Parr (BLYP) [11,14] GGAs, and the Tao-Perdew-Staroverov-Scuseria (TPSS) [29] and the newly-developed strongly-constrained and appropriate-normed (SCAN) [16] metaGGAs, as well as LSDA0. Due to the error cancellations between $E_x$ and $E_c$, LSDA is comparable to PBE in the errors of $E_{xc}$, although PBE is significantly better than LSDA for $E_x$ and $E_c$ alone. TPSS gives slightly better $E_{xc}$ than PBE does, and significantly so for $E_x$ and $E_c$ alone. Among all tested functionals, LSDA0 and SCAN are the best two for $E_{xc}$ together and for $E_x$ and $E_c$ separately. SCAN is slightly better than LSDA0, which is likely due to the fact that SCAN satisfies more exact constraints than LSDA0 does for 1- and 2-electron systems[16]. The remarkable performance of LSDA0 on the 2-electron Hooke's atoms demonstrates again that the exchange-correlation energies, and even the exchange energies alone, are quite local. On the other hand, the unsatisfactory performance of LSDA and the other functionals (PBE and TPSS) based on it implies that the dependence on $r_s$ is different here from that in the UEG. We note that BLYP is even worse than LSDA for the exchange-correlation energies, with large errors contributed by LYP to the correlation part.

The above systems used for construction and testing are those without multiple nuclear centers. Moving to multi-center systems, Table I shows that LSDA0 gives only -0.2% error in $E_{xc}$ when the hydrogen molecule ion ($H_2^+$) is at its equilibrium with R=2 Bohr for the ground gerade state, where the electron localizes between the two protons as illustrated in Fig. 3 (a). This shows again the locality of the 1- and 2-electron densities at least at the equilibrium of multi-center systems. Of course, the constructed LSDA0 cannot and should not be able to describe the stretched bond situation, where the exchange-correlation hole is highly delocalized. For example, when the bond is stretched to R=4 Bohr, where the electron delocalizes between the two protons, as shown in Fig. 3(b),

LSDA0 predicts too low energy by 5.0%, meaning the exchange hole of LSDA0 is too deep as expected. (LSDA instead gives an accurate energy here). However, we argue that LSDA0 is a better starting point for modeling the nonlocality of 1- and 2-electron systems than other traditional functionals, e.g., LSDA and the PBE and BLYP GGAs.

We can also apply LSDA0 and the other functionals to excited one-electron systems, where the exact exchange-correlation energy is naturally defined as $-U[n]$. This is related to the self-interaction correction in the Perdew-Zunger scheme[4,30]. Table 1 shows that LSDA0 yields an error of -10.9% for the excited ungerade state of $H_2^+$ at R=2 Bohr, considerably worse than that for the ground state at the same bond length. If we compare the electron densities of the gerade state at R=4 Bohr and the ungerade state at R=2 Bohr, the latter has a more delocalized electron density than the former due to the nodal plane at the center between the two protons. This explains the worse performance of LSDA0 for the excited state at the same bond length, e.g., at R=2 Bohr. Among the functionals, LSDA is surprisingly the best, while LSDA0 is considerably better than PBE and TPSS because PBE and TPSS see the great inhomogeneity of the electron density around the nodal plane and deepen their model exchange-correlation holes too much. (At a node, the reduced density gradient $s = |\nabla n|/[2(3\pi^2)^{1/3} n^{4/3}]$ diverges.)

Similarly, Table 2 shows the performance of the functionals for the ground and excited states of the hydrogen atom. Here, LSDA0 is much better than PBE and TPSS, as expected from the argument about the nodal planes, and is considerably better than LSDA as well.

We now define a measure of the localization of the exact exchange hole around its electron by

$$L = \frac{E_x^{exact}}{1.174 E_x^{LSDA}} \leq 1. \tag{4}$$

Fig. 2(d) shows that the exchange hole is highly localized in the Hooke's atoms, while, in Tables 1 and 2, $L$ decreases as the hole delocalizes in $H_2^+$ (consistent with the density plots of Fig. 3) and the hydrogen atom. When one electron is shared among N mutually-distant protons, $L = \frac{(\frac{1}{N})^2}{(\frac{1}{N})^{4/3}} = (\frac{1}{N})^{2/3}$. When L is close to 1, a good semilocal functional will be accurate for exchange and more accurate for exchange and correlation together.

We have shown the near-locality of exchange and correlation for 1- and 2-electron systems by invoking the optimal lower bounds on their exchange-correlation energies. These and other exact constraints lead us to the LSDA0 approximation for such systems. We have shown that LSDA0 achieves remarkable accuracy for 1- and 2-electron ground-state densities, apart from a necessary failure in situations where electrons are shared over stretched bonds. For such situations, and also for noded or excited-state 1-electron densities, a fully-nonlocal Perdew-Zunger self-interaction correction is recommended.

Thus small finite systems have their own version (LSDA0) of the local spin density approximation. That fact could have a significant impact on DFT, as UEG-based LSDA does. Indeed, this insight is already integrated into the SCAN meta-GGA without and with long-range van der Waals correction. Together with the knowledge of locality of an orbital defined here, a judicious application of the Perdew-Zunger self-interaction correction to these general-purpose functionals might solve many of the remaining problems of these approximations (after allowing for spin symmetry breaking when it is energetically preferred).

**Acknowledgments:** This work was mainly supported by NSF under grant DMR-1305135 (JS, JPP, and ZY). Calculations by HP was supported as part of the Center for the Computational Design of Functional Layered Materials, an Energy Frontier Research Center funded by the U.S. Department of Energy (DOE), Office of Science, Basic Energy Sciences (BES), under Award #  DE-SC0012575. JPP thanks D.G. Truhlar for pointing out the work of Loos and Gill.

**References**

[1] W. Kohn and L.J. Sham, Phys. Rev. 140, A1133-A1138 (1965).

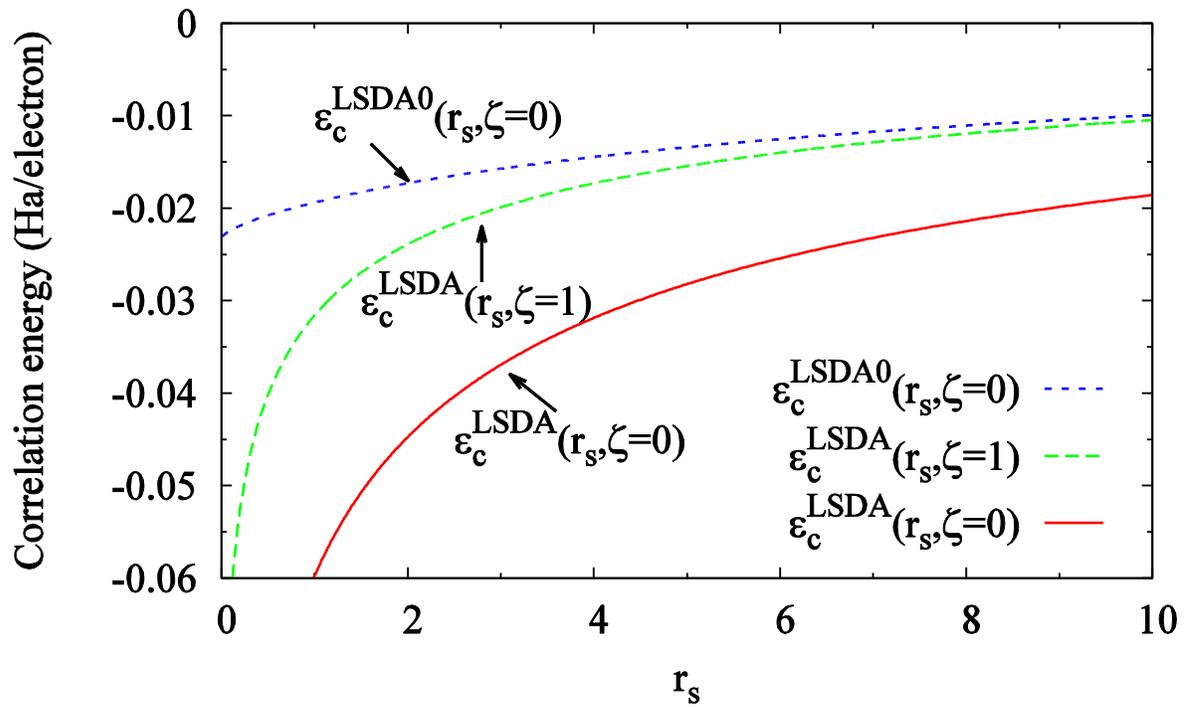

Figure 1. Correlation energy per electron for LSDA and LSDA0. Note that $\varepsilon_C^{LSDA0}(r_s, |\zeta| = 1) = 0$, since there is no correlation in a one-electron system.

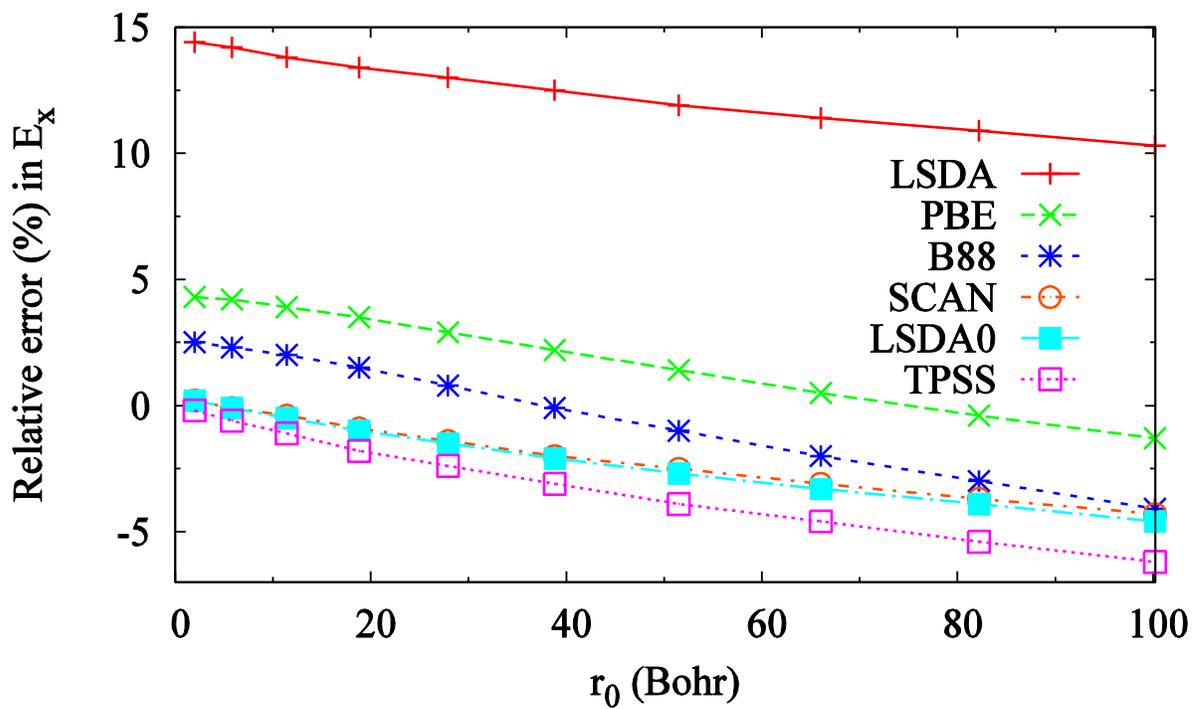

(a)

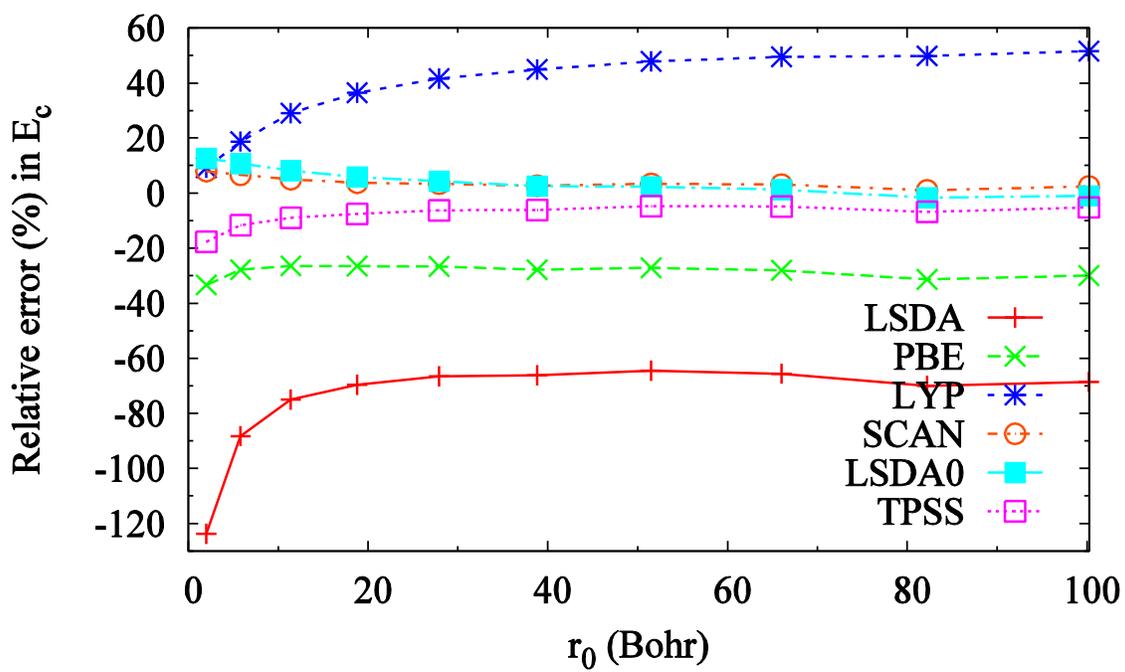

(b)

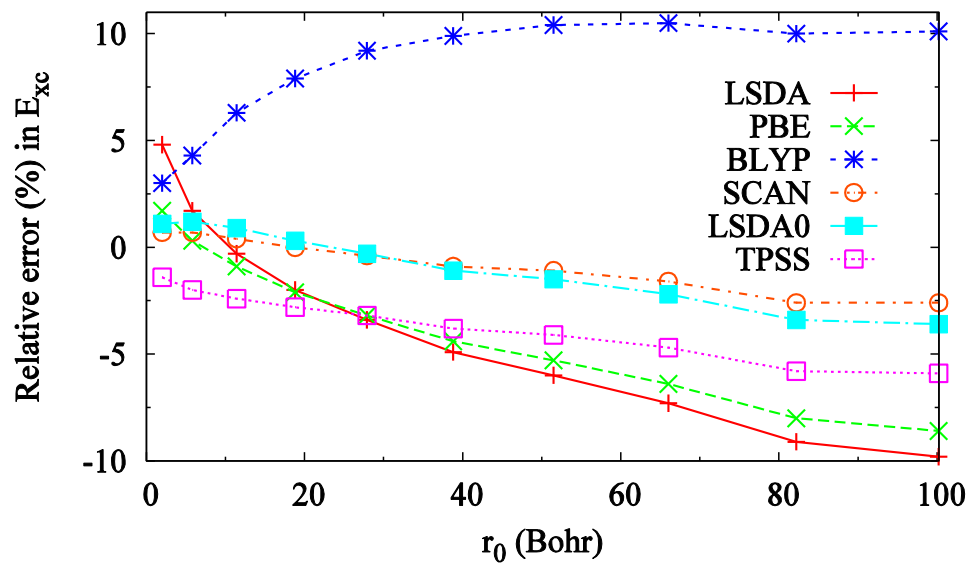

(c)

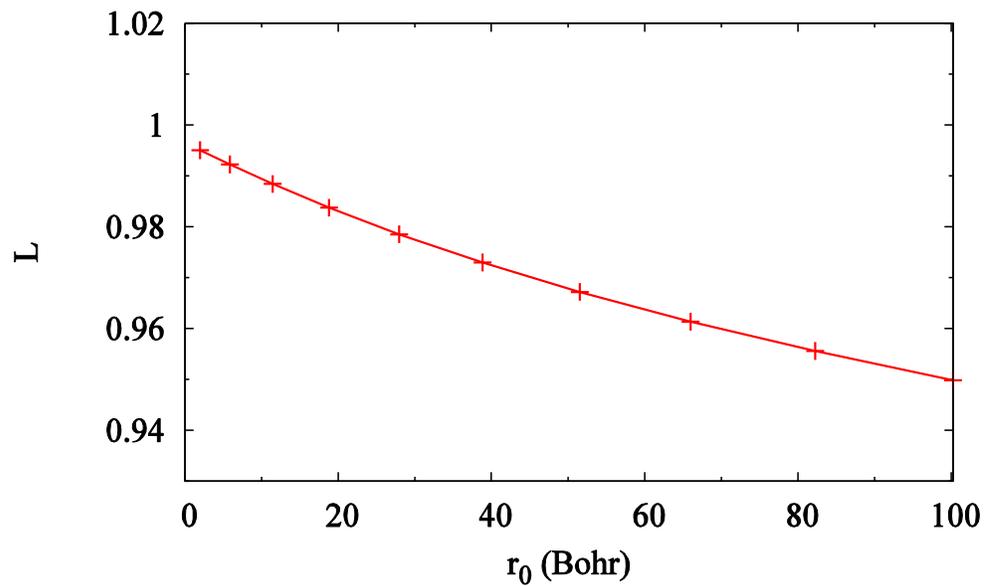

(d)

Figure 2. (a-c) Relative errors in exchange and correlation energies from different functionals, for the Hooke's atom at different classical distances between electrons, $r_0 = (\omega^2/2)^{-1/3}$, with $\omega$ the frequency of the isotropic harmonic potential. LSDA0 is close to and sometimes obscures SCAN. (d) Locality of orbitals of the Hooke's atoms.

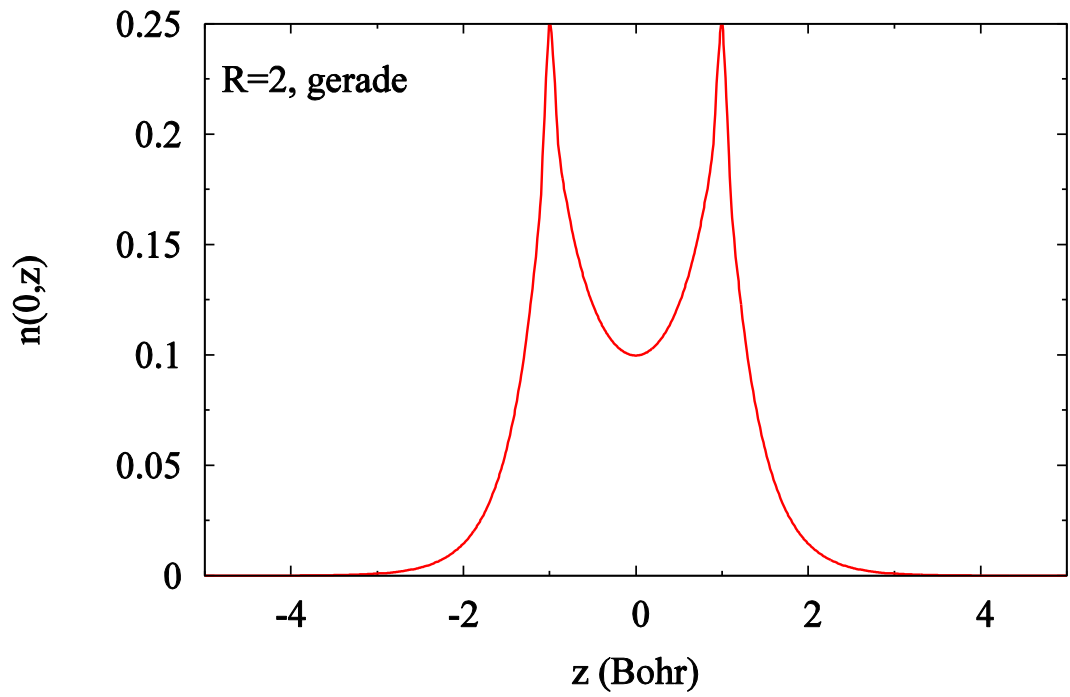

(a)

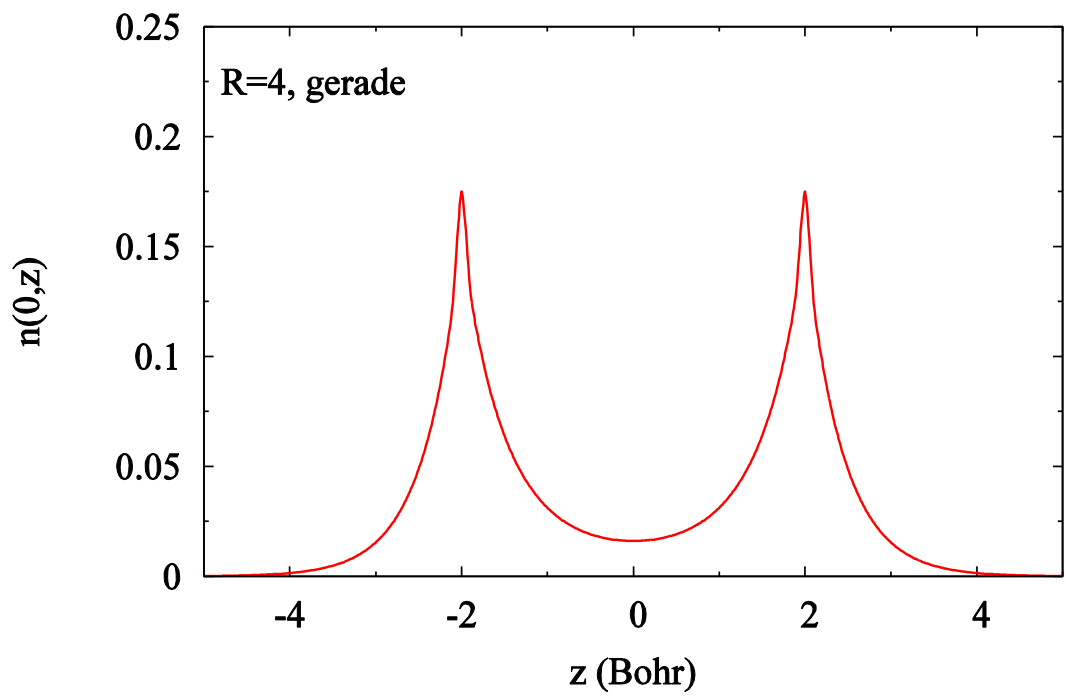

(b)

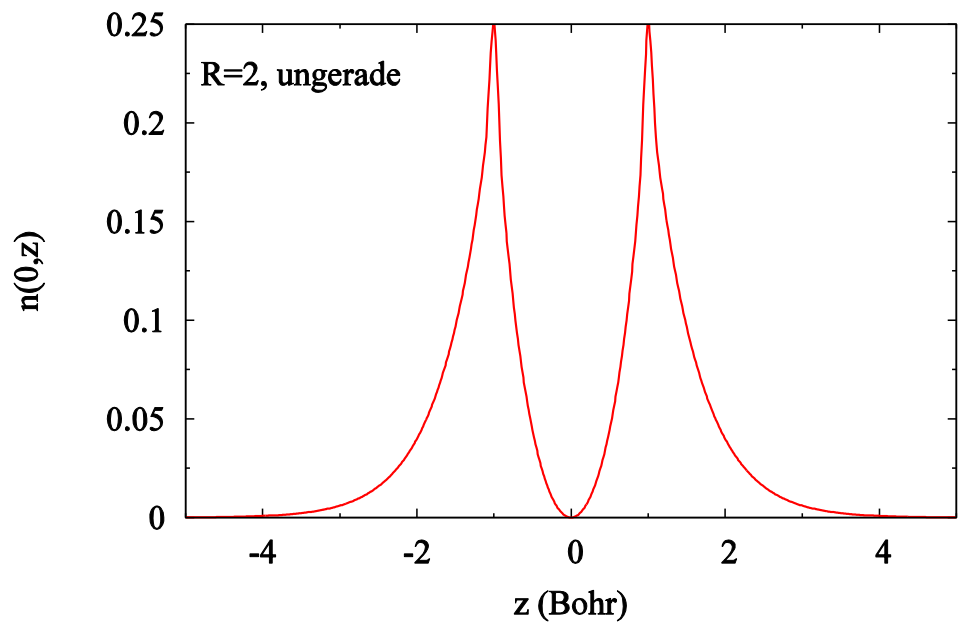

(c)

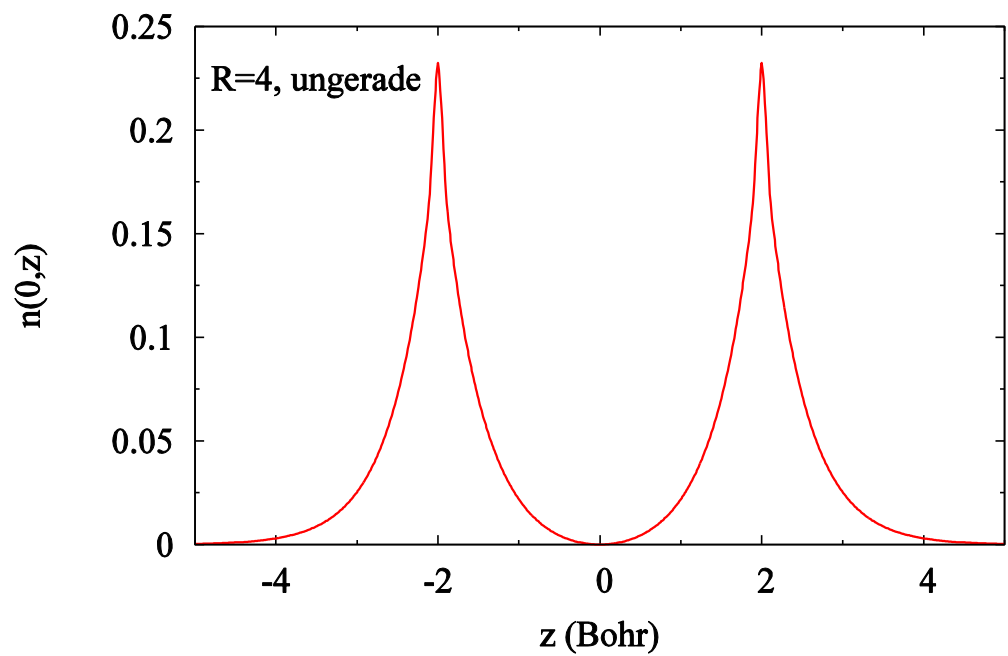

(d)

Figure 3. Electron densities along the nuclear axis of the ground-state (gerade) and excited (ungerade) $H_2^+$ at different bond lengths R (unit: Bohr). All quantities in atomic units. The equilibrium bond length is 2 Bohr.

Table 1. Relative errors (%) of different functionals for the exchange-correlation energies of the ground-state (gerade) and excited (ungerade) H2+ at different bond lengths. Note the exchange-correlation energy of the 1-electron systems is defined as minus the Hartree electrostatic interaction of the density with itself ($-U[n]$), given in Hartree units in the 3$^{rd}$ column. $L$ measures the locality of orbitals.

|  | R | -U ($E_{xc}$) | L | LSDA | PBE | TPSS | SCAN | LSDA0 |
|---|---|---|---|---|---|---|---|---|
|  | 1 | -0.37421 | 0.927 | 8.2 | 0.9 | -1.4 | 0.5 | 0.5 |
|  | 2 | -0.28935 | 0.911 | 6.5 | -0.5 | -2.2 | -0.1 | -0.2 |
|  | 3 | -0.23685 | 0.888 | 4.1 | -3.1 | -4.3 | -1.8 | -2.0 |
|  | 4 | -0.20243 | 0.857 | 0.6 | -7.1 | -7.8 | -4.8 | -5.0 |
| gerade | 5 | -0.17972 | 0.820 | -3.8 | -12.4 | -12.7 | -9.0 | -9.3 |
|  | 1 | -0.18568 | 0.824 | -3.4 | -13.2 | -13.6 | -8.5 | -9.0 |
|  | 2 | -0.20551 | 0.815 | -4.6 | -15.0 | -15.6 | -10.4 | -10.9 |
|  | 3 | -0.20434 | 0.797 | -6.9 | -17.6 | -18.1 | -12.9 | -13.5 |
|  | 4 | -0.19362 | 0.777 | -9.6 | -20.5 | -20.8 | -15.6 | -16.2 |
| ungerade | 5 | -0.18191 | 0.758 | -12.4 | -23.5 | -23.7 | -18.4 | -18.9 |

Table 2. Relative errors (%) of different functionals for the exchange-correlation energies of the ground-state and excited hydrogen atom. (n, l, m) are the quantum numbers of the states. The unit is Hartree in the 4$^{th}$ column. $L$ measures the locality of orbitals.

| Quantum number | | | -U ($E_{xc}$) | L | LSDA | PBE | TPSS | SCAN | LSDA0 |
|---|---|---|---|---|---|---|---|---|---|
| n | l | m | | | | | | | |
| 1 | 0 | 0 | -0.31250 | 0.917 | 7.1 | 0.2 | 0.0 | 0.0 | 0.0 |
| 2 | 0 | 0 | -0.07520 | 0.802 | -6.2 | -14.7 | -10.3 | -5.7 | -6.4 |
| 2 | 1 | 0 | -0.09785 | 0.794 | -7.3 | -14.8 | -11.9 | -8.8 | -9.3 |
| 3 | 0 | 0 | -0.03320 | 0.742 | -14.8 | -24.1 | -16.6 | -8.1 | -9.5 |
| 3 | 1 | 0 | -0.03881 | 0.700 | -21.6 | -31.1 | -24.2 | -16.4 | -17.7 |
| 3 | 2 | 0 | -0.04609 | 0.722 | -18.0 | -27.0 | -21.1 | -14.1 | -15.2 |
| 4 | 0 | 0 | -0.01864 | 0.703 | -21.2 | -31.1 | -21.5 | -9.4 | -11.5 |
| 4 | 1 | 0 | -0.02106 | 0.656 | -29.8 | -40.2 | -30.7 | -19.1 | -21.1 |
| 4 | 2 | 0 | -0.02282 | 0.648 | -31.4 | -42.5 | -33.6 | -21.2 | -23.3 |
| 4 | 3 | 0 | -0.02680 | 0.676 | -26.0 | -36.3 | -28.3 | -17.3 | -19.2 |